\documentclass[aps,prd,twocolumn,nofootinbib,amsmath,amssymb,showpacs]{revtex4-1}

\usepackage{graphics}
\usepackage[usenames,dvipsnames]{color}
\usepackage{hyperref}

\usepackage{lipsum}

\begin{document}
 
\title{Lepton-rich cold quark matter}

\author{J. C. {\sc Jim\'enez}}
\author{E. S. {\sc Fraga}}

\affiliation{Instituto de F\'\i sica, Universidade Federal do Rio de Janeiro,
Caixa Postal 68528, 21941-972, Rio de Janeiro, RJ, Brazil}


\begin{abstract}
We explore protoneutron star matter in the framework of cold and dense QCD using the state-of-the-art perturbative equation of state including neutrinos fixed by a lepton fraction that is appropriate for this environment. Furthermore, we calculate the modifications in the lepton-rich equation of state showing that stable strange quark matter has a more restricted parameter space.

\end{abstract}

\pacs{25.75.Nq, 11.10.Wx, 12.39.Fe, 64.60.Q-}

\maketitle

\section{Introduction}

Compact stars are natural astrophysical laboratories to investigate the nuclear strong interaction under extreme conditions \cite{Glendenning:2000}, as spectacularly illustrated by the detection of gravitational waves coming from a binary neutron star merger \cite{TheLIGOScientific:2017qsa} and the possibilities it brings to probe properties of neutron star interiors \cite{Andersson:2009yt}.

The stellar structure and the many possible phases that could be found in neutron stars stem from the equation of state (EoS) for strongly-interacting matter combined with powerful gravitational effects. Unfortunately,  the EoS is poorly known at densities of the order of the saturation density, $n_{0} \approx 0.16~$fm$^{-3}$, and above. This sector in the parameter space of quantum chromodynamics (QCD) is still not accessible to lattice simulations due to the sign problem \cite{deForcrand:2010ys}. However, one can still resort on cold and dense perturbative QCD (pQCD) \cite{kapusta-gale,Laine:2016hma}, which provides a controllable approximation even though its region of validity corresponds to much higher baryon densities. The state-of-the-art EoS within this approach was obtained in Ref. \cite{Kurkela:2009gj}, and goes well beyond the MIT bag model as it was shown in Ref. \cite{Fraga:2013qra}.

In this work, we are interested in studying the early stage of life of neutron stars, i.e., protoneutron stars (PNS). Matter in the interior of PNS is hot and lepton rich, so that one has to incorporate a non-zero fraction of trapped neutrinos in the framework \cite{Fraga:2015xha,Jimenez:2017fax}. Here, we explore PNS matter using the state-of-the-art perturbative equation of state for cold and dense quark matter, constraining it to take into account the presence of a fixed lepton fraction in which both electrons and neutrinos are included. Trapped neutrinos modify the parameter space of the cold pQCD equation of state. In addition to computing the new features in the lepton-rich equation of state due to the presence of trapped neutrinos, we also show that stable strange quark matter is less favorable in this environment. For more details, see Ref. \cite{Jimenez:2017fax}.

This work is organized as follows. In Sec. \ref{sec:QM} we briefly review some of the properties of the state-of-the-art lepton-poor EoS from pQCD. In Sec. \ref{sec:lQM} we discuss pQCD matter in the presence of electrons and trapped neutrinos to compute lepton-rich thermodynamic observables. In Sec. \ref{sec:lSQM} we study the allowed parameter space for stable strange quark matter. Finally, Sec. \ref{sec:conclusion} contains our summary.

\section{Lepton-poor quark matter}
	\label{sec:QM}

The pressure for cold and dense QCD matter in the framework of pQCD with massless quarks was first computed to order $\mathcal{O}(\alpha^{2}_{s})$ many years ago \cite{Freedman:1976ub,Baluni:1977ms} (see also Ref. \cite{Toimela:1984xy}). Later it was recomputed with a modern definition of the running coupling and used to model the nonideality in the EoS \cite{Fraga:2001id} (see also Ref. \cite{Blaizot:2000fc}), also for massless quarks. Quark mass effects have been studied in early days in Refs. \cite{Freedman:1977gz,Farhi:1984qu} in the context of quark stars \cite{Itoh:1970uw} and strange matter, respectively. Later, such effects have been described in the $\overline{\rm MS}$ scheme in Ref. \cite{Fraga:2004gz} to $\mathcal{O}(\alpha_{s})$, where it became clear that quark mass effects and its associate running through the renormalization group equations can be significant in the physics of compact stars. The state-of-the-art pQCD EoS for cold quark matter, including renormalization group effects up to $\mathcal{O}(\alpha^{2}_{s})$ in the strange quark mass and strong coupling, was obtained by Kurkela $\textit{et al.}$ \cite{Kurkela:2009gj} (see also Ref. \cite{Kurkela:2010yk} for direct comparison with astrophysical observations). 

\begin{figure}[ht]
	\begin{center}
	\resizebox*{!}{5.5cm}{\includegraphics{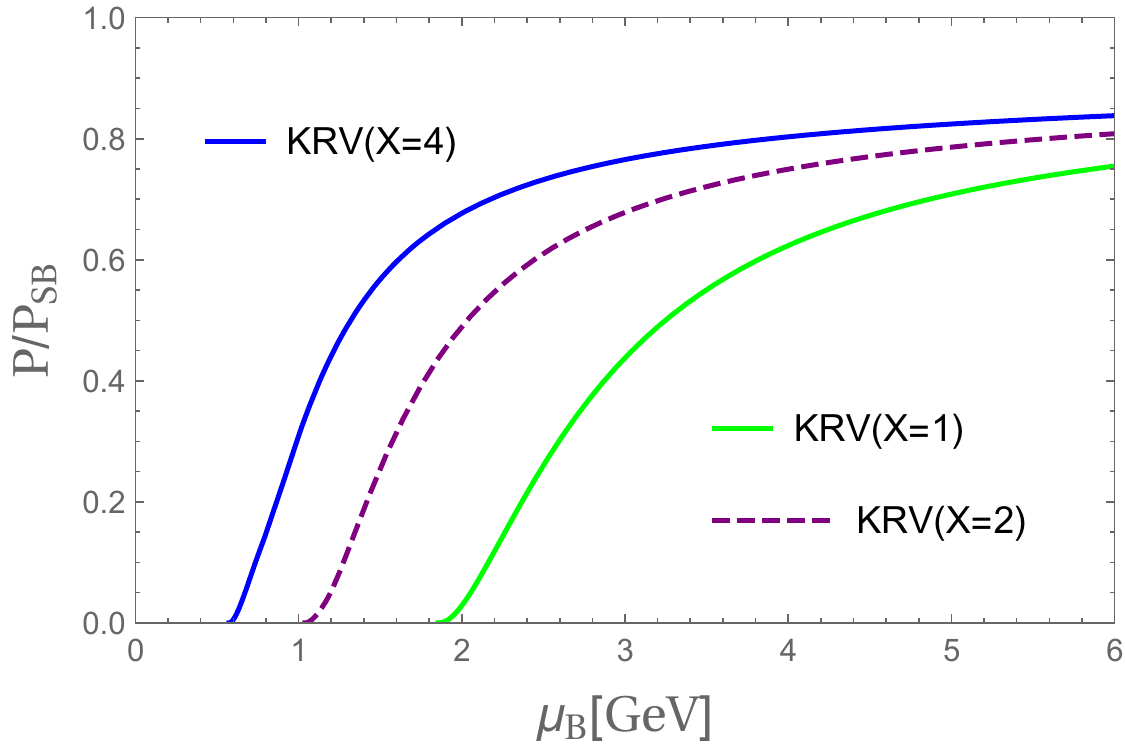}}
      \end{center}    
      \caption{ \label{fig:KRVcold}Total pressure of a gas of up, down and massive strange quarks plus electrons up to three-loops \cite{Kurkela:2009gj} normalized by the Stefan-Boltzmann free pressure for different values of $X$. Local charge neutrality and $\beta$-equilibrium were implemented.}
    \end{figure}

Since the QCD thermodynamic potential is calculated perturbatively, its truncation generates an unknown scale, $\bar{\Lambda}$, associated with the subtraction point for renormalization. For $T= m_{f}=0$, $\bar{\Lambda}$ is proportional to the quark chemical potential, $\mu$, and this freedom generates an uncertainty band. We adopt the standard fiducial scale $\bar{\Lambda}=(2/3)\mu_{B}$, where $\mu_{B}$ is the baryon chemical potential. 
For the strange quark mass we choose $m_{s}(2\rm ~GeV,N _{f}=2+1)=92 \rm ~MeV$ \cite{Aoki:2016frl} and, for the strong coupling constant, $\alpha_{s}(1.5$GeV$,N_{f}=3)=0.336$, which allows us to fix the renormalization point in the $\overline{\rm MS}$ scheme to $\Lambda_{\overline{\rm MS}}=315^{+18}_{-12}$ MeV \cite{Bazavov:2014soa}. For convenience, we define the dimensionless renormalization scale to be $X \equiv 3\bar{\Lambda}/\mu_{B}$. The behavior of the pressure obtained in Ref. \cite{Kurkela:2009gj}, which we refer to as KRV from now on, as a function of $\mu_{B}$ is illustrated in Fig. \ref{fig:KRVcold}.

\section{Lepton-rich quark matter}
	\label{sec:lQM}
    
 In order to explore the first moments of life of neutron stars as protoneutron stars formed at the early postbounce stage of core collapse supernovae, one has to consider an EoS that is rich in leptons. In particular, one has to include trapped neutrinos in the pQCD framework. In the following, we set a widely accepted value for lepton fraction in PNS matter, i.e.,  $Y_{L}={0.4}$ \cite{Burrows:1986me,Pons:1998mm}, which forces us to introduce a chemical potential for the neutrinos, $\mu_{\nu}$. 

As usual, we define the total quark number density as
\begin{equation}
n=n_{u}(\mu_{u},X)+n_{d}(\mu_{d},X)+n_{s}(\mu_{s},X) \, ,
\end{equation}
where each quark density depends on its respective chemical potential and on the renormalization scale $X$. The baryon number density is defined as $n_{B}=n/3$. 
Since we aim to study the properties of compact stars, one has to impose charge neutrality and lepton fraction conservation. Doing it locally:
      \begin{equation}
	\frac{2}{3}n_{u}-\frac{1}{3}n_{d}-\frac{1}{3}n_{s}=n_{e} \, ,
	  \label{Eq2}
      \end{equation}
      \begin{equation}
	\frac{n_{e}+n_{\nu}}{n_{B}}=Y_{L}=0.4 \, ,
      \end{equation}
while the weak interaction equilibrium conditions imply
      \begin{equation}
	\mu_{d}+\mu_{\nu}=\mu_{u}+\mu_{e},
      \end{equation}
      \begin{equation}
	\mu_{d}=\mu_{s}\equiv{\mu} \, .
		\label{Eq5}
      \end{equation}
Here, $\mu_{u}$, $\mu_{d}$, $\mu_{s}$, $\mu_{e}$ and $\mu_{\nu}$ are the chemical potentials of the up, down and strange quarks, the electron and the electron neutrino. The latter are introduced as degenerate Fermi gas contributions.

    \begin{figure}[ht]
	\begin{center}
	\resizebox*{!}{5.5cm}{\includegraphics{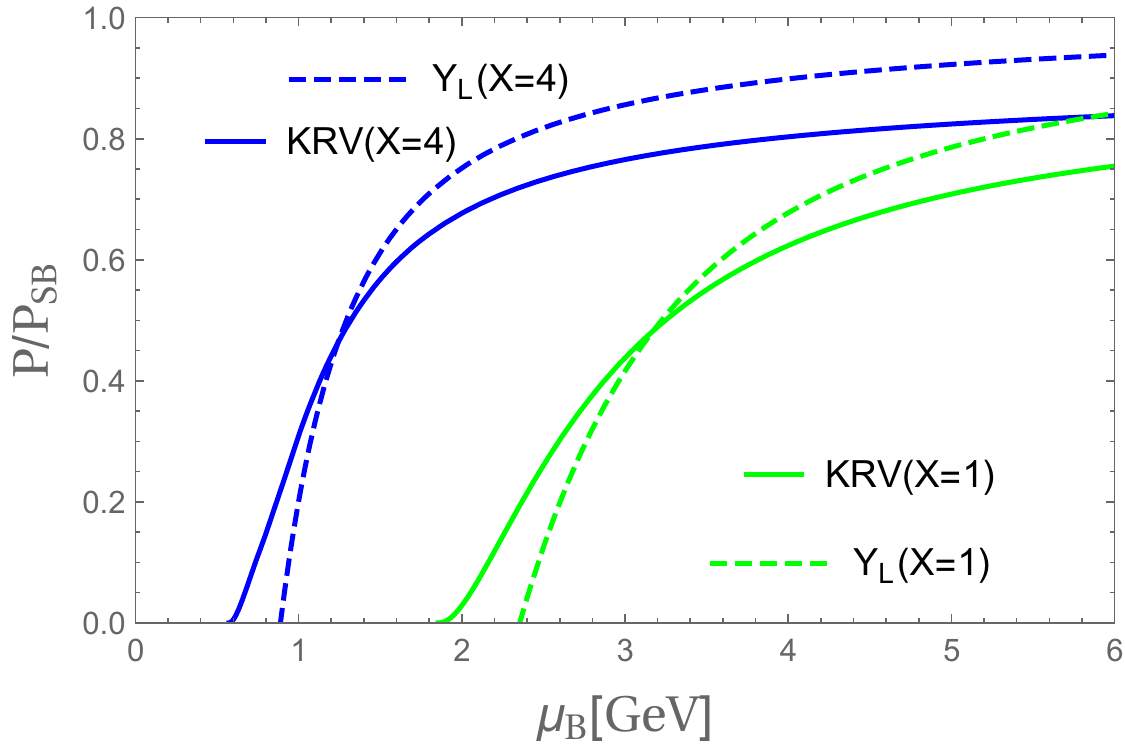}}
      \end{center}    
      \caption{\label{fig:KRVneutrino}Total pressure of quarks and leptons for a fixed lepton fraction ($Y_{L}=0.4$) in dashed lines for the band $X\in[1, 4]$. In solid lines we show the lepton-poor case (KRV).}
    \end{figure}

Using the constraints above, we can write all quark and lepton chemical potentials in terms of the strange quark chemical potential, $\mu_{s}\equiv{\mu}$, only. Following Kurkela $\textit{et al.}$ \cite{Kurkela:2009gj}, we use the quark and lepton number densities as the building blocks from which one can construct the thermodynamic variables, in our case the total pressure, demanding thermodynamic consistency at each step of the calculation and preserving terms up to $\mathcal{O}(\alpha^{2}_{s})$. 

\begin{figure}[ht]
	\begin{center}
	\resizebox*{!}{5.5cm}{\includegraphics{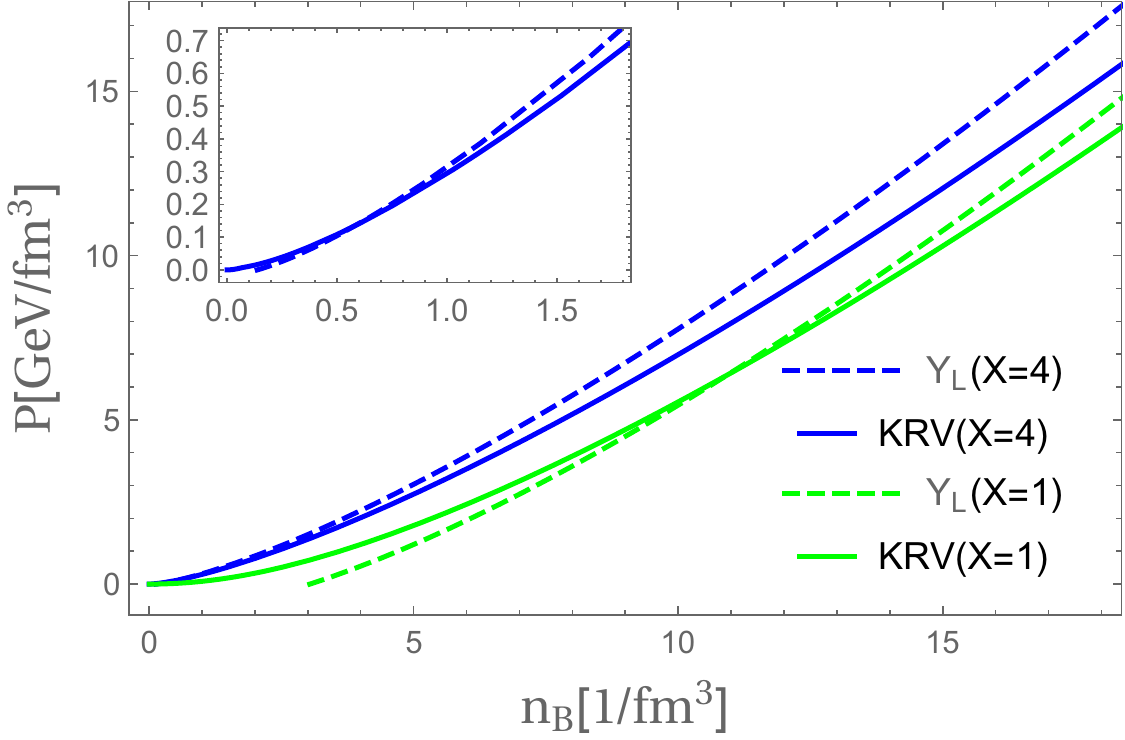}}
      \end{center}    
      \caption{\label{fig:PnBYl} Total pressure of quarks and leptons for a fixed lepton fraction, $Y_{L}(X)$, and the lepton-poor case, KRV(X), as a function of the baryon number density $n_{B}$. 
      }.
    \end{figure}

When some quark density flavor $n_{f}(\mu_{f},X)$ becomes negative below a given chemical potential flavor, $\mu_{f}<\mu_f^{0}(X)$, we set it to $n_{f}\equiv{0}$. Integrating the number densities from their minimal value $\mu_f^{0}(X)$ to some arbitrary strange quark chemical potential $\mu$ and taking into account Eqs. (\ref{Eq2})--(\ref{Eq5}), we obtain the total pressure for lepton-rich quark matter as follows:
\begin{eqnarray}
P(\mu,X)=\int^{\mu}_{\mu_{0}(X)}d\bar{\mu}
\left[ n_{u}\left(1+\frac{d\mu_{\nu}}{d\mu_{s}}-\frac{d\mu_{e}}{d\mu_{s}}\right) \right.\nonumber \\
\left. +~ n_{d}+n_{s}+
n_{e}\frac{d\mu_{e}}{d\mu_{s}}+n_{\nu}\frac{d\mu_{\nu}}{d\mu_{s}} \right] \, .
\end{eqnarray}
We express the pressure as a function of the baryon chemical potential, $P=P(\mu_{B})$, where $\mu_{B}=\mu_{u}+\mu_{d}+\mu_{s}$. In Fig. \ref{fig:KRVneutrino} one can see how the cold quark matter EoS (KRV) is modified by the presence of trapped neutrinos ($Y_{L}$) for different values of the renomalization scale $X$. Notice that, at high $\mu_{B}$, since $m_{s}(X)$ tends to be constant \cite{Fraga:2004gz,Kurkela:2009gj} and $\alpha_{s}(X)$ is non-zero (unless we are at asymptotically high densities \cite{Kurkela:2009gj}), the total pressure of quarks and leptons will increase faster, in contrast to the lepton-poor case. However, at low $\mu_{B}$, the lepton-poor total pressure appears to be higher (see Ref. \cite{Jimenez:2017fax} for a discussion). To clarify this issue, we show, in Fig. \ref{fig:PnBYl}, the total pressure of quarks and leptons as a function of $n_{B}$.

  \section{Lepton-rich stable strange quark matter}
  	\label{sec:lSQM}

Bodmer \cite{Bodmer:1971we} and Witten \cite{Witten:1984rs} investigated in different contexts a system formed by massless up, down and strange quarks. If they have at zero pressure a energy per baryon
  \begin{equation}
{E}/{A}\leq{0.93}\rm GeV \, ,
\label{EoverA}
\end{equation}
i.e., lower than the most stable nuclei $\rm Fe^{56}$, one would find configurations of absolutely $\textit{stable strange quark matter}$ (SQM) as the true ground state of hadronic matter in the vacuum.



We explore the criterion above using the thermodynamic potential discussed in Sec. \ref{sec:lQM}.
 To do that, we use the Hugenholtz-Van Hove theorem \cite{Hugenholtz:1958zz} generalized to a system with many components \cite{R.C.Nayak:2011hyj}. It requires only the quark and lepton densities and chemical potentials as input, giving the following energy per baryon:
	\begin{equation}
	\frac{E(\mu_{s},X)}{A}=\frac{n_{u}}{n_{B}}(\mu_{\nu}-\mu_{e})+3\mu_{s} \, ,
	\end{equation}
where we implicitly assumed that all the quantities on the rhs of the equation above are functions of the strange chemical potential $\mu_{s}$ and renormalization scale $X$.

\begin{figure}[ht]
	\begin{center}
	\resizebox*{!}{5.5cm}{\includegraphics{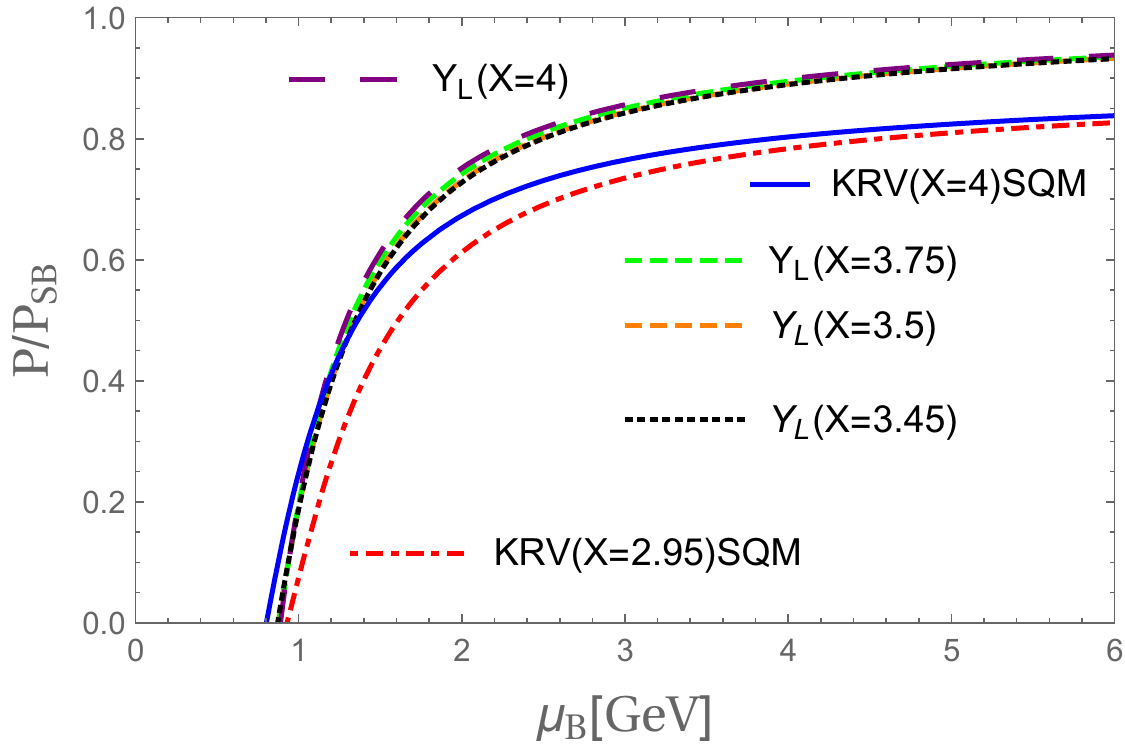}}
      \end{center}    
      \caption{\label{fig:SQMcomparison} Total normalized pressure for quarks and leptons with $Y_{L}=0.4$ allowing the SQM hypothesis ($Y_{L}$, in dashed lines). For comparison, we show also the pressure for lepton-poor strange matter ($\rm KRV[X]SQM$, in solid and dot-dashed lines).}
    \end{figure}

Constraining the values of $X$ such that $\mu_{s}$ and $n_{s}$ are not zero $\textit{and}$ satisfy Eq. (\ref{EoverA}) we obtain, for the cold case, $X\in[2.95, 4]$, and for the lepton-rich case $X\in[3.45, 4]$, as it can be seen in Fig. \ref{fig:SQMcomparison}. Even if the parameter space of $X$ is not radically modified when trapped neutrinos are included, one can notice from Fig. \ref{fig:SQMcomparison} that the band for $X$ tends to shrink to $\mu_{B}\in[0.86, 0.88]\rm GeV$ for vanishing pressure (as compared to $\mu_{B}\in[0.803, 0.93]\rm GeV$ in the cold case). Figure \ref{fig:SQMcomparison} also indicates that lepton-rich strange quark matter becomes essentially $\textit{independent}$ of the renormalization scale $X$.

One can infer from this that the presence of leptons fixed to some non-trivial value makes the SQM hypothesis $\textit{less}$ favorable, i.e., the stability windows of critical densities with vanishing pressure is narrower. A similar behavior was observed in Ref. \cite{Dexheimer:2013czv}, where the authors also included finite-temperature contributions in different quark models. This leaves us with $X\in[1, 3.44]$ for lepton-rich quark matter having as ground state hadronic matter in the vacuum.

\section{Summary}
  \label{sec:conclusion}

In this work we have explored protoneutron star matter, i.e., stellar matter formed after a core-collapse supernova explosion which after a minute produces a regular neutron star, using the state-of-the-art perturbative equation of state for cold and dense QCD matter in the presence of a fixed lepton fraction in which both electrons and neutrinos are included. Finite-temperature effects on the equation of state can be neglected since they have a minor effect in the PNS scenario we have in mind where densities are high enough compared to thermal effects. Even if the presence of neutrinos does not modify appreciably the EoS at low densities compared with other effective models for lepton-rich quark matter, for some values of the renormalization scale $X$ their presence significantly increases the pressure as one goes to higher densities, still within the region relevant for the physics of PNS, making the EoS stiffer.

Modifications in the equation of state due to the presence of trapped neutrinos were investigated showing that the stable strange quark matter hypothesis is less favorable in this environment, i.e., the parameter space for the formation of strange quark matter with neutrinos decreases considerably. 

\begin{acknowledgments}
This work was supported by CNPq and FAPERJ, being also part of the project INCT-FNA Process No. 464898/2014-5.
\end{acknowledgments}


\end{document}